\documentclass[10pt,twoside]{article}
\usepackage{rsfs} 
\usepackage{graphicx}
\usepackage{KYBER2E,amssymb,latexsym,amscd,amsmath,a4,epsfig}

\newfont{\mi}{cmti9}
\newfont{\m}{cmr8}
\newfont{\ms}{cmsl8}
\newfont{\autor}{cmcsc10}
\newcommand{\Section}[1]{\section{#1}\setcounter{equation}{0}}

\newtheorem{theorem}{Theorem} [section]
\newtheorem {lemma}[theorem]{Lemma}

\newtheorem {corollary}[theorem]{Corollary}

\newtheorem {example}[theorem]{Example}
\newtheorem {examples}[theorem]{Examples}

\newtheorem {remark}[theorem]{Remark}
\newtheorem {remarks}[theorem]{Remarks}
\newcommand {\ar}{\rightarrow}
\newcommand{\beq}{\begin{equation}}
\newcommand{\eeq}{\end{equation}}
\newcommand{\Leq}[1]{\label{#1}\end{equation}}
\newcommand{\beqn}{\begin{eqnarray}}
\newcommand{\eeqn}{\end{eqnarray}}
\newcommand{\beqno}{\begin{eqnarray*}}
\newcommand{\eeqno}{\end{eqnarray*}}
\renewcommand {\l}{\left}
\newcommand {\ri}{\right}
\newcommand{\bem}{\l(\! \begin{array}}
\newcommand{\eem}{\end{array}\!\ri)}
\newcommand{\bsm}{\left(\begin{smallmatrix}} 
\newcommand{\esm}{\end{smallmatrix}\right)}  
\newcommand{\qmbox}[1]{\quad\mbox{#1}\quad}
\newcommand {\eh}{{\textstyle \frac{1}{2}}}
 
\newcommand {\bN}{{\mathbb N}}
\newcommand {\bR}{{\mathbb R}}

\newcommand{\cF}{{\mathcal F}}
\newcommand{\cI}{{\mathcal I}}
\newcommand{\cM}{{\mathcal M}}
\newcommand{\cP}{{\mathcal P}}
\newcommand{\tp}{{\tilde p}}
\newcommand{\supp}{{\rm supp}}
\newcommand{\GCD}{{\rm GCD}}
\newcommand{\ov}{\overline}
\newcommand{\es}{\emptyset}

\begin{document}

\rule[3mm]{128mm}{0mm}
\vspace*{-16mm}

{\footnotesize K\,Y\,B\,E\,R\,N\,E\,T\,I\,K\,A\, ---
\,V\,O\,L\,U\,M\,E\, {\it 4\,0\,} (\,2\,0\,0\,4\,)\,,\,
N\,U\,M\,B\,E\,R\, 1\,,\,\  P\,A\,G\,E\,S\, \,x\,x\,x --
x\,x\,x}\\ \rule[3mm]{128mm}{0.2mm}

\vspace*{11mm}

{\large\bf \noindent Maximizing Multi-Information}


\vspace*{23mm}

\small

Stochastic interdependence of a 
probablility distribution on a product space is measured
by its Kullback-Leibler distance
from the exponential family of product distributions (called multi-information). 
Here we investigate low-dimensional exponential families that contain the maximizers of stochastic interdependence
in their closure. 

Based on a detailed 
description of the structure of probablility distributions with globally maximal multi-information we 
obtain our main result: The exponential family of pure pair-interactions contains all global maximizers of the 
multi-information in its closure.

\smallskip\par
\noindent {\sl Keywords:}\,
\begin{minipage}[t]{112mm}
Multi-information, exponential family, relative entropy,
pair-interaction, infomax principle, Boltzmann machine, neural networks.
\end{minipage} 

\normalsize


\

\Section{\bf Introduction}
The starting point of this article is a geometric interpretation of the 
interdependence\footnote{Throughout the paper we use the term 
{\em interdependence\/} 
to indicate stochastic dependence among units, as opposed to dependence
of general random variables.} of
stochastic units. In order to illustrate the basic idea, we consider two units with the configuration sets
$\Omega_1 = \Omega_2 = \{0,1\}$. The configuration set of the whole system is just the Cartesian product
$\Omega_1 \times \Omega_2 = \{(0,0),(1,0),(0,1),(1,1)\}$. The set of probability 
distributions ({\em states}) is a 
three-dimensional simplex $\overline{\mathcal P}(\Omega_1\times \Omega_2)$
with the four extreme points $\delta_{(\omega_1,\omega_2)}$, 
$\omega_1,\omega_2 \in \{0,1\}$ (Dirac measures). The two units are independent with respect to 
$p \in \overline{\mathcal P}(\Omega_1 \times \Omega_2)$ iff 
\begin{equation} \label{factor}
   p(\omega_1,\omega_2) \; = \; p_1(\omega_1)\,p_2(\omega_2)\quad 
   \mbox{for all} \quad (\omega_1,\omega_2) \in \Omega_1 \times \Omega_2.  
\end{equation}
The set of {\em factorizable\/} distributions (\ref{factor}) is a two-dimensional manifold ${\mathcal F}$.
Figure~1 shows the simplex $\overline{\mathcal P}(\Omega_1\times \Omega_2)$ and 
its submanifold ${\mathcal F}$. 

\begin{figure}
\begin{center}
\setlength{\unitlength}{1cm}
\begin{picture}(7,7)
\put(0,0){\epsfbox{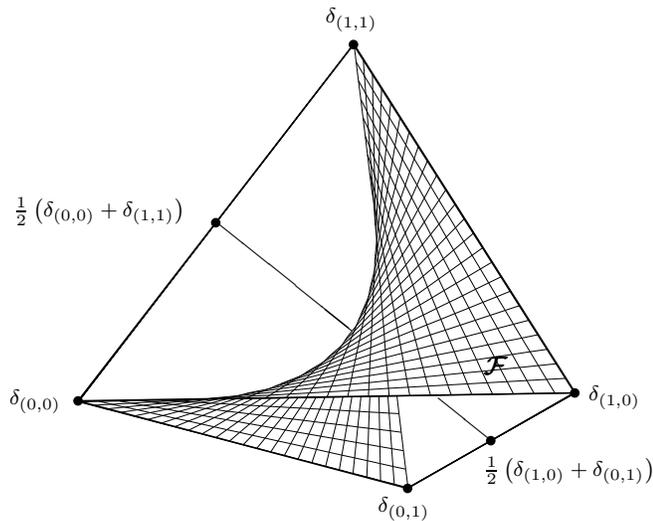}}
\put(3.4,6.28){\footnotesize $\delta_{(1,1)}$}
\put(-0.75,3.7){\footnotesize 
$\frac{1}{2}\left(\delta_{(0,0)} + \delta_{(1,1)}\right)$}
\put(5.5,0.25){\footnotesize 
$\frac{1}{2}\left(\delta_{(1,0)} + \delta_{(0,1)}\right)$}
\put(6.9,1.25){\footnotesize $\delta_{(1,0)}$}
\put(4.1,-0.23){\footnotesize $\delta_{(0, 1)}$}
\put(-0.8,1.2){\footnotesize $\delta_{(0,0)}$}
\put(5.5,1.6){$\boldsymbol{\mathcal F}$}
\put(1.95,3.625){\line(5,-4){1.8}}
\put(4.9,1.29){\line(5,-4){0.7}}
\end{picture}
\end{center}

\caption{The exponential family $\cF$ in the simplex of probability distributions.}
\end{figure}
\vspace{1cm}

Given an arbitrary probability distribution $p$, we quantify the interdependence of the two units 
with respect to $p$ by its Kullback-Leibler distance from the set ${\mathcal F}$. In our two-unit case,
this distance is nothing but the well known mutual information, which has been introduced by
Shannon \cite{Sh} as a fundamental quantity that provides a measure of the capacity 
of a communication channel. 

Motivated by so-called {\em Infomax principles\/} within the field 
of neural networks \cite{Li,TSE}, one of us has investigated maximizers of the
interdependence \cite{Ay1,Ay2} of stochastic units. In our two-unit example, these are the distributions
\[
  \eh\left( \delta_{(0,0)} + \delta_{(1,1)} \right), \quad \mbox{and}\quad
  \eh\left( \delta_{(1,0)} + \delta_{(0,1)} \right) \qquad (\mbox{see Figure 1}).
\]
This article continues that work by analyzing the structure of maximizers of stochastic
interdependence. In particular, this leads to some answers to the   
question on the existence and the structure of a natural low dimensional manifold that contains all 
maximizers of the stochastic interdependence
(see \cite{Ay1}, 3.4 (ii) and \cite{Ay2}, 4.2.3). We will prove that the exponential family 
of pure pair-interactions contains the global maximizers of multi-information in its closure. In our 
example of two binary units this exponential family is given by the convex hull of the two 
maximizers $\eh\left( \delta_{(0,0)} + \delta_{(1,1)} \right)$ and 
$\eh\left( \delta_{(1,0)} + \delta_{(0,1)} \right)$ shown in Figure 1. 

In physics, pair interactions 
are considered as fundamental mechanisms that underly most theories. Within the field of 
neural networks, the physical concept of pair-interactions is used to model the synaptic interactions
of neurons.  

\Section{\bf Notation}
Let $\Omega$ be a nonempty and finite set. In the corresponding real vector space ${\Bbb R}^\Omega$, we have
the canonical basis $e_\omega$, $\omega \in \Omega$, which induces the natural scalar product 
$\langle \cdot , \cdot \rangle$.\\
The set of probability distributions
on $\Omega$ is denoted by $\overline{\mathcal P}(\Omega)$:
\[
   \overline{\mathcal P}(\Omega) \; := \; 
   \left\{ p = \big(p(\omega)\big)_{\omega \in \Omega}  \in {\Bbb R}^\Omega \; : \; 
   \mbox{$p(\omega) \geq 0$ for all $\omega$, and $\sum_{\omega \in \Omega} p(\omega) = 1$} \right\}.
\]
For a probability distribution $p$, we consider its support 
${\rm supp}\, p := \{ \omega \in \Omega \; : \; p(\omega) > 0 \}$. The strictly positive distributions
${\mathcal P}(\Omega)$ have maximal support $\Omega$:
\[
  {\mathcal P}(\Omega) \; := \; \{p\in\overline{{\mathcal P}}(\Omega) \; : \; {\rm supp} \, p =\Omega\}.
\] 
Note that $\overline{{\mathcal P}}(\Omega) $ is the closure of ${\mathcal P}(\Omega)$.
For every vector $X = (X(\omega))_{\omega \in \Omega} \in {\Bbb R}^\Omega$, 
we consider the corresponding {\em Gibbs measure\/}:
\[
      \exp(X)\in\cP(\Omega)\qmbox{,} 
      \exp(X)(\omega) := \frac{e^{X(\omega)}}{\sum_{\omega' \in \Omega} e^{X(\omega')}}.
\]
The image 
$\exp({\mathcal T})$ of a linear (or more generally affine) subspace 
${\mathcal T}$ of ${\Bbb R}^\Omega$ with respect to the map
$X \mapsto \exp(X)$ is called {\em exponential family\/} 
({\em induced by ${\mathcal T}$\/}).\\  
In this article, we are mainly interested in the ``distance'' of probability distributions from a given 
exponential family ${\mathcal E}$. More precisely, we use the {\em Kullback-Leibler divergence}
or {\em relative entropy}
$D: \overline{\mathcal P}(\Omega) \times \overline{\mathcal P}(\Omega) \rightarrow 
[0,\infty)\cup\{\infty\}$,
\[
   (p,q) \; \mapsto \; 
   D(p \, \| \, q) := 
   \left\{
      \begin{array}{c@{,\quad}l}
         \sum_{\omega\in {\rm supp}\,p} p(\omega) \ln \frac{p(\omega)}{q(\omega)} & 
         \mbox{if ${\rm supp}\,p \subset {\rm supp}\, q$}, \\
         \infty   & \mbox{otherwise}
      \end{array}
   \right.,
\]
to define the continuous\footnote{See Lemma 4.2 of \cite{Ay1} for a proof.}  
function $D_{\mathcal E}: \overline{\mathcal P}(\Omega) 
\rightarrow {\Bbb R}_+$, 
\[
  p \; \mapsto \;  D_{\mathcal E}(p) \; := \; \inf_{q \in {\mathcal E}} D(p \, \| \, q). 
\] 

For $k\in \bN$ we denote the set $\{1,\ldots,k\}$ by $[k]$.  
%
\Section{\bf Sufficiency of Low-Dimensional Exponential Families for the Maximization of 
Multi-Information} 
We consider
the set $V:=[N] = \{1,\dots,N\}$ of $N \geq 2$ {\em units}, and corresponding sets 
$\Omega_i$, $i\in [N]$, of 
{\em configurations\/}. The number $|\Omega_i|$ of configurations of a unit $i$ 
is denoted by $n_i$. 
Without restriction of generality 
we assume
\[
  2 \; \leq \; n_1 \; \leq \; n_2 \; \leq \; \cdots \; \leq \; n_N.
\]
For a subsystem $A \subseteq [N]$, the set of configurations on $A$ is given by the 
product $\Omega_A := \times_{i \in A} \Omega_i$. 
One has the natural restriction 
\[
  X_A: \Omega_V \rightarrow \Omega_A\qmbox{,}(\omega_i)_{i\in [N]} \mapsto 
  (\omega_i)_{i\in A},
\] 
which induces the projection 
\[
  \overline{\mathcal P}(\Omega_V) \rightarrow \overline{\mathcal P}(\Omega_A)
  \qmbox{,} p \mapsto p_A,
\] 
where $p_A$ denotes the image measure of $p$ with respect to the variable $X_A$. For  
$i \in [N]$ we write $p_i$ instead of $p_{\{i\}}$.\\ 
A probability distribution 
$p \in \overline{\mathcal P}(\Omega_V)$ is called {\em factorizable\/} if it satisfies
\[
   p(\omega_1,\dots,\omega_N) \; = \; 
   p_1(\omega_1)\cdot\ldots\cdot p_N(\omega_N) \quad \mbox{for all} \quad  
   (\omega_1,\dots,\omega_N) \in \Omega_V.
\] 
The set ${\mathcal F}$ of strictly positive and factorizable probability
distributions on $\Omega_V$ is an exponential family in ${\mathcal P}(\Omega_V)$ with
\[
     {\rm dim}\,{\mathcal F} \; = \; \sum_{i =1}^N (n_i - 1) .
\]
Now let us consider the function $D_{\mathcal F}$, which measures the distance from ${\mathcal F}$. 
We have $D_\cF(p)=0$ if and only if $p\in\overline{\cP}(\Omega_V)$ is factorizable. Thus, this distance function can be interpreted as a measure that    
quantifies the stochastic interdependence of the units in $[N]$. The following entropic 
representation of $D_{\mathcal F}$ is well known (see \cite{Am3}):
\[
     I_p(X_1,\dots,X_N) \; := \; D_{\mathcal F}(p) \; = \; \sum_{i=1}^N H_p(X_i) - 
     H_p(X_1,\dots,X_N).
\]  
Here, the $H_p(X_i)$'s denote the marginal entropies and 
$H_p(X_1,\dots,X_N)$ is the global entropy. This measure of
stochastic interdependence of the units, which is called
{\em multi-information}, is a generalization of the mutual information (see example in the introduction).\\
This article deals with the problem of finding natural low-dimensional exponential families that contain the maximizers of the multi-information in their closure. 
To this end we first consider a result 
on maximizers of the distance from an {\em arbitrary} exponential family \cite{Ay1},
in the improved form obtained in 
\cite{MA}:
\medskip

\noindent
{\bf Prop.\ 3  of \cite{MA}. }\label{thm:2.2}
Let ${\mathcal E}$ be an exponential family in ${\mathcal P}(\Omega)$
with dimension $d$. Then there exists an exponential family 
${\mathcal E}^\ast$, ${\mathcal E} \subset {\mathcal E}^\ast$, with dimension 
less than or equal to $3 d + 2 $ such that 
the topological closure of 
${\mathcal E}^\ast$ contains all local 
maximizers of $D_{\mathcal E}$.
\medskip

This theorem is quite general, and is based on the observation that maximizers of the 
information divergence $D_{\mathcal E}$ have a reduced cardinality of their support, 
which is controlled by the
dimension $d$ of ${\mathcal E}$. 
%
%
The direct application of Prop. 3 of \cite{MA}
to the exponential family 
${\mathcal F}$ leads to the following statements on the local maximizers of the
multi-information $I(X_1,\dots,X_N) = D_{\mathcal F}$:
\medskip


\begin{corollary} \label{thm:2.4}
There exists an exponential family ${\mathcal F}^\ast$ with
\[
   {\rm dim}\,{\mathcal F}^\ast \; \leq \; 3 \sum_{i =1}^N (n_i - 1) + 2 \; \leq \; 
   3 N (n_N-1) + 2
\]
that contains all local maximizers of $I(X_1,\dots,X_N)$ in its topological closure.\\
In particular, in the binary case $n_i = 2$ 
for all $i$, ${\rm dim}\,{\mathcal F}^\ast \leq 3 N + 2 $. 
\end{corollary}
In all such statements about exponential families over product spaces 
one should keep in mind, that the dimension of the exponential family 
$\cP(\Omega_V)$ itself is of exponential growth in the number $N=|V|$
of units. So any exponential subfamily which is of polynomial growth in $N$
is of large codimension.

Our main goal is now the following. Knowing about the existence of such low-dimensional exponential 
families ${\mathcal F}^\ast$, we want to analyze the relation between them and exponential families 
given by interaction structures between the $N$ units. 

More precisely, this article deals with the problem whether 
one can find low-dimensional exponential families ${\mathcal F}^\ast$
like in the Corollary \ref{thm:2.4} that are at the same time given by a
low order of interaction. Before going into the details, we state an informal version of the main result 
of the paper (using terminology from statistical physics):
\medskip

{\bf Informal Version of Theorem \ref{thm:main}:} 
{\em If the cardinalities $n_1,\dots,n_N$ fulfill an inequality 
(see Theorem \ref{thm:4.4}), the exponential family of pure 
pair-interactions (that is, pair-interactions without any external field) 
is sufficient for generating all global maximizers
of the multi-information.\/}
\medskip

Let us have a closer look on this result for the binary case. In this case, the 
exponential family of pure pair-interaction has dimension $N-1$,  
which is stronger than Corollary \ref{thm:2.4}. 
More important, the {\em pair} interactions form an explicit 
low dimensional exponential family that appears in many 
models in physics and biology 
(the units being called particles respectively neurons, the interactions fields 
resp.\ dendrites). \\
In Section 5, we will provide a rigorous formulation of our main result and prove it. This will be based 
on results concerning the structure of global maximizers of multi-information, 
which is discussed in the following Section 4.
\medskip

\Section{\bf The Structure of Global Maximizers of Multi-Information} 
\subsection{General Structure}
Obviously, the maximal value of $I(X_1,\dots,X_N)$ is bounded as
\[
   I_p(X_1,\dots,X_N) \; = \; 
   \sum_{i=1}^N H_p(X_i) \; - \; H_p(X_1,\dots,X_N) \; \leq \; \sum_{i=1}^{N} \ln(n_i).
\]
In fact, 
it turns out that in contrast to the quantum setting (see Remark \ref{rmk:3.1.5} below), 
this upper bound is never reached.
The following lemma gives an upper bound that is sharp in 
many interesting as well as important 
cases.
\medskip

\begin{lemma} \label{lemma1}
Let $p$ be a probability distribution on 
$\Omega_V = \Omega_1 \times \cdots \times \Omega_N$. Then: 
\begin{equation} \label{abschmax}
   I_p(X_1,\dots,X_N) \; \leq \; \sum_{i = 1}^{N-1} \ln (n_i).
\end{equation}
\end{lemma}
\medskip

\begin{remark} \label{rmk:3.1.5} 
With an orthonormal basis $f_1,\dots,f_n$ 
of the Hilbert space ${\Bbb C}^{n}$ we consider 
the (entangled) unit vector 
\[
    \psi \; := \; \frac{1}{\sqrt{n}} \sum_{k = 1}^n \bigotimes_{i=1}^N f_k\ \in\ 
    \bigotimes_{i=1}^N{\Bbb C}^n,
\]  
and the density operator $\rho$ defined by the orthogonal projection onto
the subspace spanned by $\psi$. In this setting, the mutual information is extended as
\[
    I(\rho) \; = \; \sum_{i=1}^NS(\rho_i) - S(\rho) \; = \; 
    {\rm tr}\big(\rho \ln(\rho)\big) - \sum_{i=1}^N {\rm tr}\big(\rho_i \ln (\rho_i)\big)
\]
where $S$ denotes von Neumann entropy, and the $\rho_i$ 
are the partial traces of $\rho$. 
As we see, this multi-information 
has the value $N \ln(n)$, which, according to Lemma \ref{lemma1}, 
is not possible within the classical setting.
\end{remark}
\medskip

In the following, we consider the set
\[\cM(\Omega_1, \dots, \Omega_N) :=
\left\{p\in\overline{\cP}(\Omega_V)\ : \  
 I_p(X_1,\dots,X_N) = \sum_{i = 1}^{N-1} \ln (n_i)\right\}\]
of probability distributions that maximize, according to  
Lemma \ref{lemma1},  in the case
$\cM(\Omega_1, \dots, \Omega_N)\neq \emptyset$
the multi-information $I(X_1,\dots,X_N)$.
Up to isomorphism, everything depends only on the cardinalities 
$n_i=|\Omega_i|$ so that we sometimes write $\cM(n_1, \dots, n_N)$
instead of $\cM(\Omega_1, \dots, \Omega_N)$.

The next theorem characterizes the probability distributions in 
$\cM(\Omega_1, \dots , \Omega_N)$.
\medskip

\begin{theorem}\label{prop1}
Let $p$ be a probability distribution on 
$\Omega_V$. Then $p \in \cM(\Omega_1, \dots , \Omega_N)$
if and only if
there exist a probability distribution $p^{(N)} \in \overline{\mathcal P}(\Omega_N)$ 
and surjective maps $\pi_i: \Omega_N \rightarrow \Omega_i$, $i=1,\dots,N-1$,  
with
\begin{equation} \label{surjektionen}
  p^{(N)}\left\{ \pi_i = \omega_i\right\}     \; = \; \frac{1}{n_i}\qquad
(\omega_i \in \Omega_i),
\end{equation}
such that for all $(\omega_1,\dots,\omega_N) \in \Omega_V$ 
\begin{equation} \label{darstellung}
   p(\omega_1,\dots,\omega_N)  \; = \; 
   \left\{
      \begin{array}{c@{,\quad}l}
         p^{(N)}(\omega_N) & \mbox{if $\omega_i = \pi_i(\omega_N)$, $i=1,\dots,N-1$}, \\
         0                 & \mbox{otherwise}.
      \end{array}
   \right.
\end{equation}
\end{theorem}
\medskip

Theorem \ref{prop1} allows us to say  precisely under which conditions 
on the unit sizes $n_i$ the theoretical 
maximum (\ref{abschmax}) of multi-information can be achieved 
(we use the shorthands $W:=2^{[N-1]}\backslash\{\emptyset\}$ and 
$n_A:=(n_i)_{i\in A}$ and denote the greatest common divisor by ${\rm GCD}$):
\begin{theorem} \label{thm:4.4}
We have
$\cM(\Omega_1,\ldots,\Omega_N)\neq\emptyset$
if and only if $n_N\geq n_{\min}$ for
\[n_{\min}=n_{\min}(n_1,\ldots, n_{N-1}):=
\sum_{A\in W}(-1)^{|A|-1}{\rm GCD}(n_A).\]
\end{theorem} 

\begin{remarks}
\begin{enumerate}
\item
In particular, $\cM(\Omega_1,\ldots,\Omega_N)\neq\emptyset$ if
\begin{enumerate}
\item
there are only $N=2$ units, or
\item
all units are identical $(n_1=\ldots=n_N)$.
\end{enumerate}
In the following Sections \ref{twounits} and \ref{identical} 
we discuss these two important examples
of Theorem \ref{thm:4.4} more precisely. 

\item
\begin{enumerate}
\item
We have the following inequalities for $n_{\min}$:
\[\max(n_1,\ldots,n_{N-1})\leq n_{\min}\leq 1+\sum_{i=1}^{N-1}(n_i-1). \]
These follow immediately from the defining relation 
$n_{\min} = |\bigcup_{i\in[N-1]}T_{n_i}|$ for 
$T_m:=\l\{\frac{i}{m}\ : \  i\in[m]\ri\}$, since 
$|T_{n_i}|=n_i$ and $1\in T_{n_i}$.

The left inequality becomes an equality iff the least common multiple
${\rm LCM}(n_{[N-1]})=n_{N-1}$ 
(still assuming that $n_{i+1}\geq n_i$), whereas
the right inequality becomes an equality iff the integers 
$n_1,\ldots,n_{N-1}$ are mutually prime.
\item
Additionally, one gets
\[n_{\min}\leq {\rm LCM}(n_{[N-1]})=:l,\]
since for all $i\in[N-1]$ the inclusion $T_{n_i}\subset T_l$ holds true.
Again we have equality iff ${\rm LCM}(n_{[N-1]})=n_{N-1}$.
\item
The global maximizers $p\in\cM(\Omega_1,\ldots,\Omega_N)$ 
of multi-information that we construct simultaneously maximize 
the mutual information of the pairs $\{i,N\}$ of units.

In the case ${\rm LCM}(n_{[N-1]})=n_{N}$ they even simultaneously maximize 
the mutual information of all pairs $\{i,j\}\subset[N]$ of units.\\
Both statements follow from direct inspection of $p$ defined in (\ref{p:c}).

\end{enumerate}
\end{enumerate}
\end{remarks}
%
%
%
%
%
%
%

\medskip

\subsection{The Case of Two Units} \label{twounits}
We now discuss the case of two units, i.e.\ $N=2$. In this case, the set 
\[
  \cM(\Omega_1,\Omega_2) \; = \;  \left\{p \in \overline{\mathcal{P}}(\Omega_1\times\Omega_2)\; : \; 
  I_p(X_1,X_2) = \ln(n_1)\right\}
\]
is non-empty and therefore consists of all global maximizers of the mutual information of the two units.
We want to describe the structure of $\cM(\Omega_1,\Omega_2)$ by stratifying it into a disjoint union 
of relatively open sets. In order to do that,
we consider for $\Omega_1^\ast := \Omega_1\cup\{0\}$ the following set of maps 
\begin{equation} \label{poset:S}
{\mathcal S} \; := \; \{\pi:\; \Omega_2 \; \to \; \Omega_1^\ast \; : \; \pi(\Omega_2)\supset\Omega_1\}.
\end{equation} 
The relation
\[
  \sigma \preceq \pi  \qquad :\Longleftrightarrow \qquad 
  \sigma^{-1}(\omega_1) \subset \pi^{-1}(\omega_1) \;\; \mbox{for all} \;\;
  \omega_1\in\Omega_1
\]
on ${\mathcal S}$ is a partial order which makes ${\mathcal S}$ a poset.
\medskip

\begin{example} \label{example:poset}{\em
For $\Omega_1=\{1,2\}$ and $\Omega_2=\{1,2,3\}$ we get a
poset ${\mathcal S}$ of 12 maps. The right graphics in Figure 2 shows the 
cover graph of the poset with vertex set ${\mathcal S}$.
On the left we show the graphs of four of these maps. We have $\sigma\preceq\pi$
if $\sigma$ is in the lower line and connected to $\pi$
in the upper line (so-called Hasse diagram).

\begin{figure}
\begin{center}
\epsfig{file=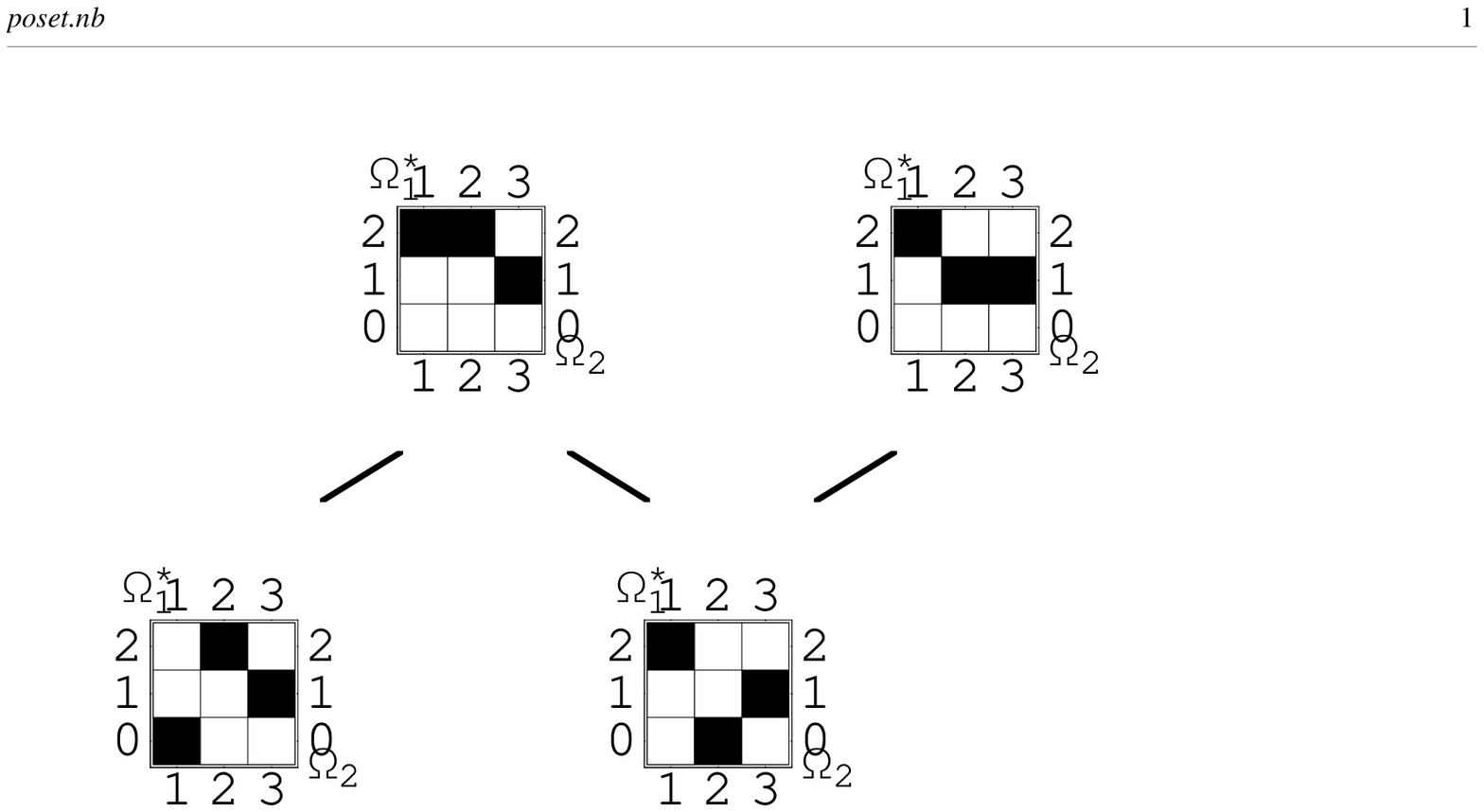,bbllx=90,bblly=475,bburx=540,bbury=715,width=9cm,clip=}
\hspace*{-15mm}
\epsfig{file=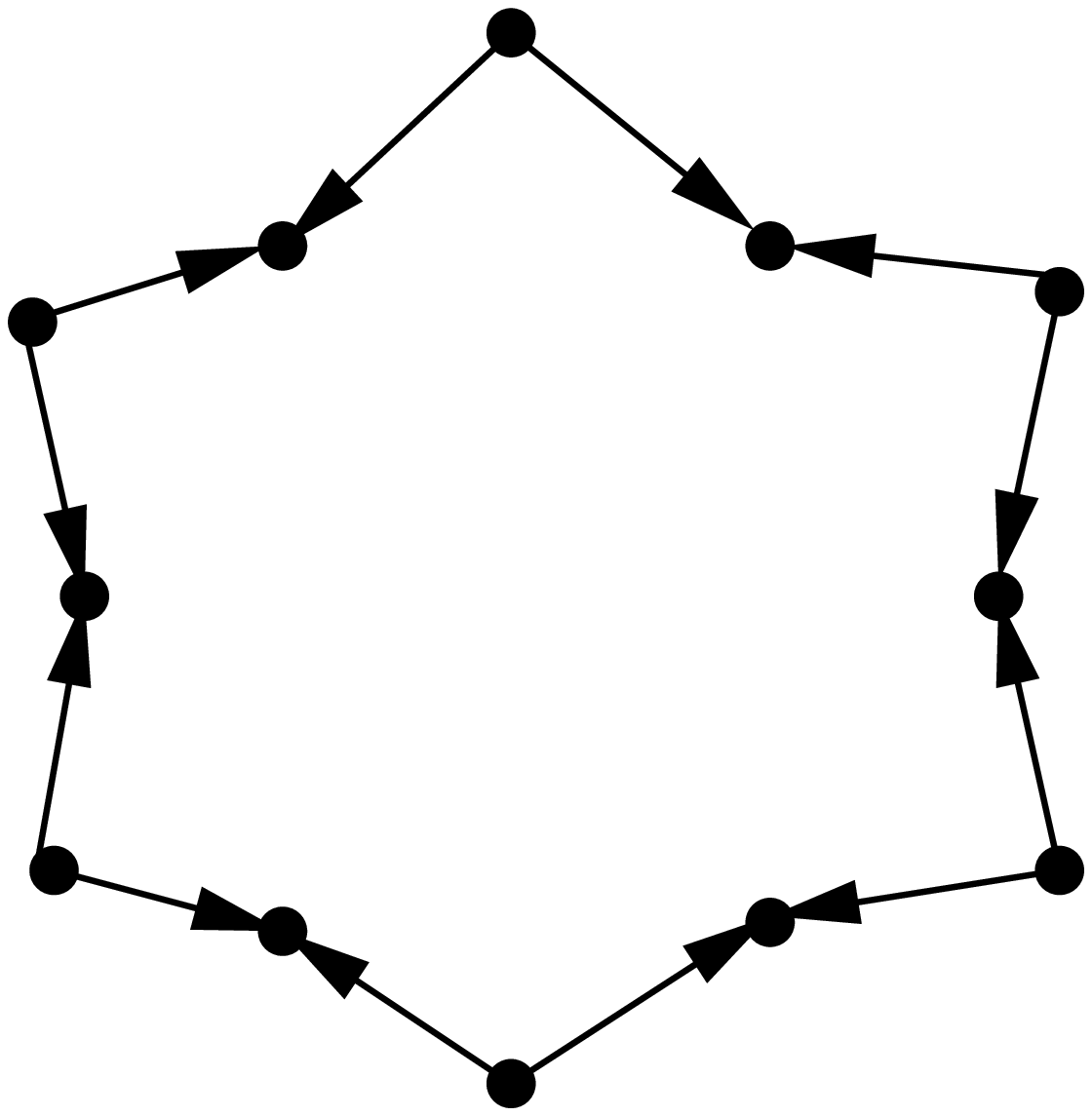,bbllx=0,bblly=-50,bburx=365,bbury=325,width=4.5cm,clip=}
\end{center}
\caption{The posets for $\Omega_1= \{1,2\}$, $\Omega_2 = \{1,2,3\}$.}
\end{figure}
}\end{example}

We call a poset {\em connected\/} iff its cover graph is connected.
\medskip

\begin{lemma}\label{lem:poset}
The poset (\ref{poset:S}) is connected if and only if $n_1 < n_2$.
\end{lemma}
\medskip

Given $\pi\in {\mathcal S}$ we consider the convex and relatively open set
\begin{eqnarray*}
\cM_\pi(\Omega_1,\Omega_2) 
& := & \Bigg\{ p \in \overline{\mathcal{P}}(\Omega_1\times\Omega_2) \; : \; 
\mbox{for all $\omega_1\in  \Omega_1$,} \\
&    & \sum_{\omega_2\in\pi^{-1}(\omega_1)}p(\omega_1,\omega_2) = \frac{1}{n_1}\quad \mbox{and} \quad
       p(\omega_1,\omega_2)>0\ \mbox{iff}\ \pi(\omega_2)=\omega_1 \Bigg\}.
\end{eqnarray*}
We denote by $S_{m,n}$ the Stirling numbers of the second kind (see for example \cite{Ai}).
\medskip

\begin{theorem} \label{thm:two:units} $ $ \\ 
{\bf (1)} The set of global maximizers of the mutual information is a disjoint union
\[
\cM(\Omega_1,\Omega_2) \; = \; {\biguplus}_{\pi\in {\mathcal S}}\cM_\pi(\Omega_1,\Omega_2)
\]
of sets $\cM_\pi(\Omega_1,\Omega_2)$.\\
{\bf (2)} These sets have dimension
\[
 \dim \cM_\pi(\Omega_1,\Omega_2) \; = \; |\pi^{-1}(\Omega_1)|-|\Omega_1|,
\]
and there are $n_1!{n_2\choose l}S_{l,n_1}$ sets $\cM_\pi(\Omega_1,\Omega_2)$ of dimension $l-n_1$.\\
{\bf (3)} The inclusion 
$\cM_\sigma(\Omega_1,\Omega_2) \subset \overline{\cM_\pi}(\Omega_1,\Omega_2)$ holds
if and only if $\sigma\preceq\pi$, 
and the set $\cM(\Omega_1,\Omega_2)$ is connected if and only if $n_1 < n_2$. 
\end{theorem}
\medskip

\begin{example} \label{example:hexagon}{\em 
Continuing Example \ref{example:poset},
for $n_1=2$ and $n_2 = 3$ the set $\cM(2,3)$ is the disjoint union of six points
and six open intervals (see Figure 3, left), combined in the form of a 
hexagon (see Figure 3, right). So $\cM(2,3)$ is homeomorphic
to $S^1$ in this case.
\medskip

\begin{figure}
\begin{center}
\epsfig{file=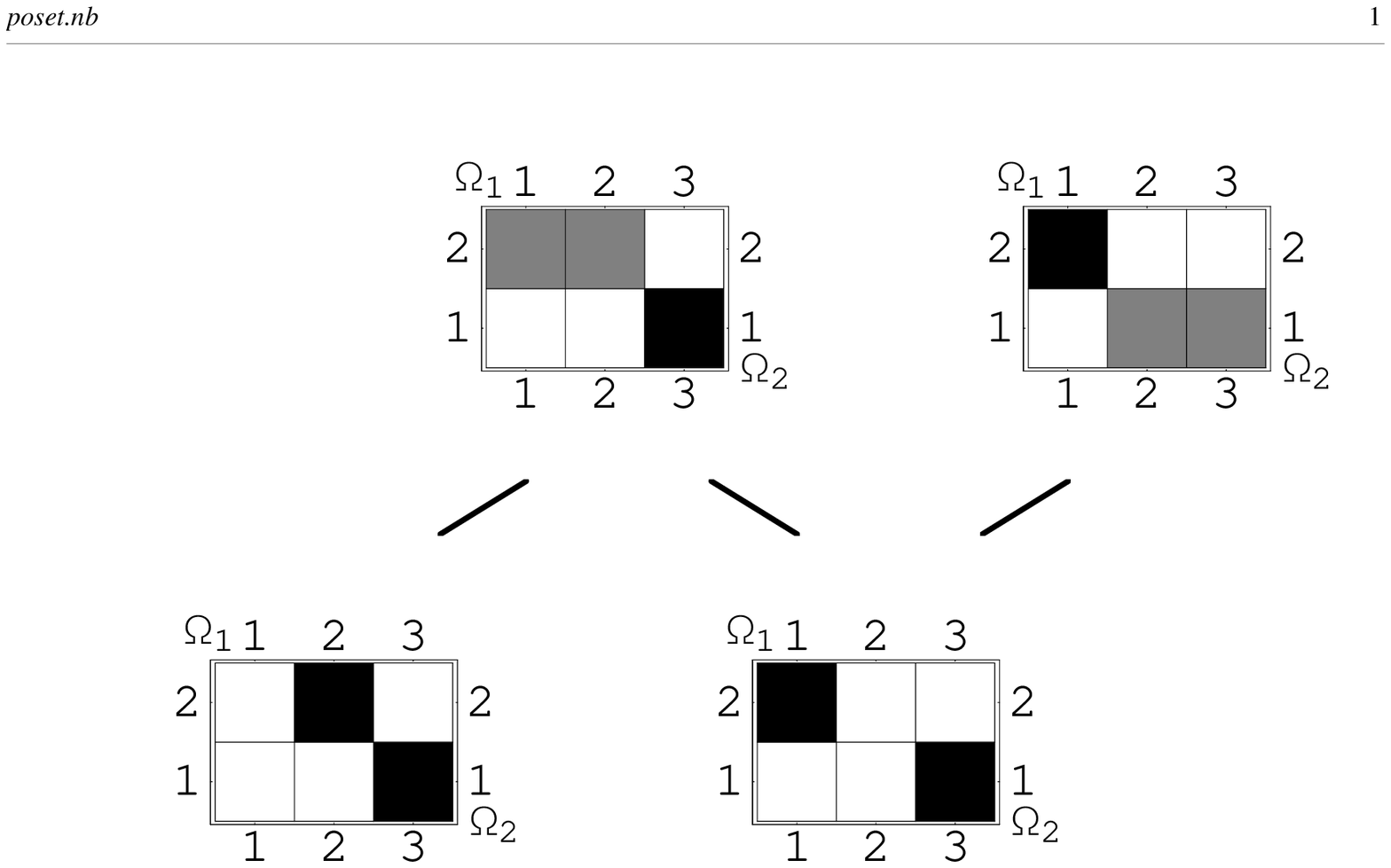,
bbllx=90,bblly=435,bburx=540,bbury=715,width=7.5cm,clip=}
\hspace*{5mm}
\epsfig{file=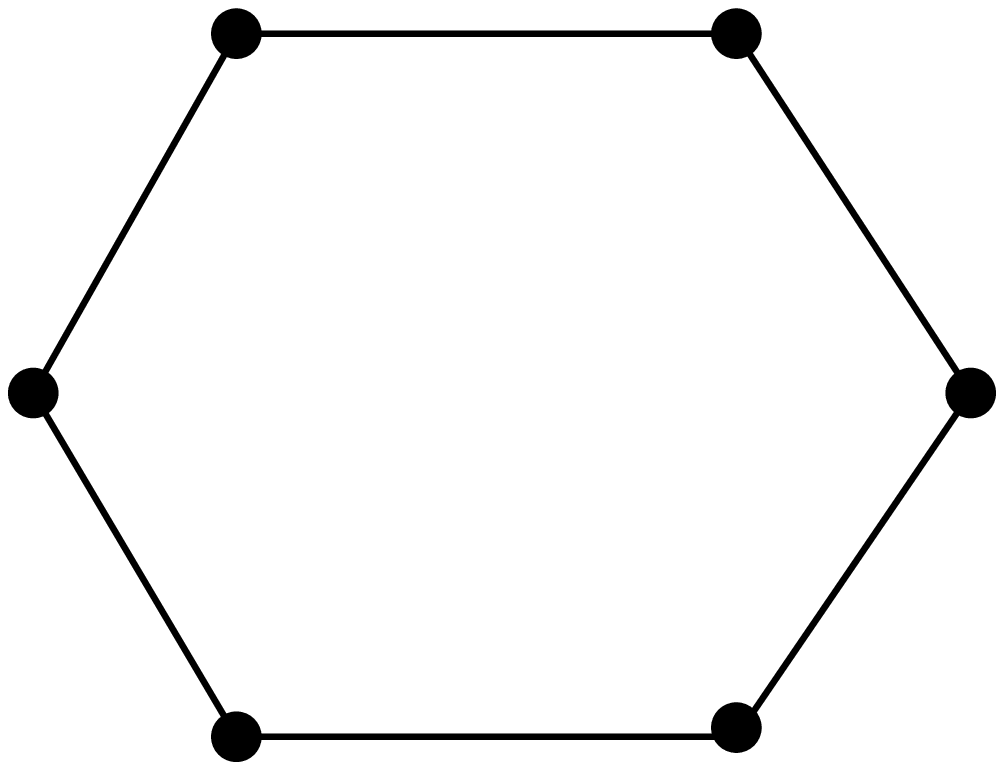,
bbllx=0,bblly=-50,bburx=325,bbury=325,width=4.5cm,clip=}
\end{center}
\caption{The structure of $\cM(2 , 3 )$.}
\end{figure}
}
\end{example} 
\bigskip

%

\subsection{The Case of $\boldsymbol{N}$ Equal Units} \label{identical}

This section deals with the important example of $N$ units with $n_1 = \dots = n_N =:n$. 
In that situation, Theorem  \ref{prop1} has the following direct implication. 
\medskip

\begin{corollary} \label{thm:same:card}
The set $\cM(\Omega_1,\dots,\Omega_N)$ consists of all probability distributions
\beqno
   \frac{1}{n} \sum_{\omega_N \in \Omega_N} 
   \delta_{(\pi_1(\omega_N),\dots,\pi_{N-1}(\omega_N),\omega_N)},  
\eeqno
where $\pi_i: \Omega_N \to \Omega_i$, $i=1,\dots,N-1$, are one-to-one mappings. 
This implies 
\begin{equation} \label{anzahl}
  |\cM(\Omega_1,\dots,\Omega_N)| \; = \; (n!)^{N-1},
\end{equation}
and for all $p \in \cM(\Omega_1,\dots,\Omega_N)$,
\beqno 
  I_p(X_1,\dots,X_N) \; = \; (N-1) \cdot \ln(n), 
\eeqno
\begin{equation} \label{neueabsch} 
  |{\rm supp}\, p| \; = \; n. 
\end{equation}
\end{corollary}
\medskip

Thus according to (\ref{anzahl}), the number of the maximizers of the multi-information grows 
exponentially in $N$. 
In particular, for binary units the set $\cM(\Omega_1,\dots,\Omega_N)$ has $2^{N-1}$ elements.
In view of this fact, it is interesting that
according to Corollary \ref{thm:2.4} there is an exponential family of dimension $\leq 3N +2$ that 
approximates all these global maximizers of the multi-information. This bound can even be improved. 
Although it is not our main goal to do that we close this subsection by an interesting $N$-independent 
upper bound, which implies that for $N$ binary units there exists an 
exponential family with dimension less than or equal to $5$ that approximates all $2^{N-1}$ 
elements of $\cM(\Omega_1,\dots,\Omega_N)$. 
\medskip

\begin{theorem} \label{thm:starkered} 
There exists an exponential family 
with dimension less than or equal to $(n^2+3n)/2$ that contains 
$\cM(\Omega_1, \dots, \Omega_N)$ in its closure.
\end{theorem} 
\medskip

This exponential family, however, is based on multibody interactions (in terms of statistical mechanics)
between the units $i\in [N]$.
\Section{\bf Sufficiency of Low-Order Interaction for the  
Maximization of Multi-Information}\label{gentheo}
Given a subset $A\subseteq [N] = \{1,\ldots,N\}$,
we decompose $\omega\in\Omega_V$ in the form $\omega \;=\;
(\omega_A,\omega_{[N]\setminus A})$ with 
$\omega_A\in\Omega_A$, $\omega_{[N]\setminus A}\in\Omega_{[N] \setminus A}$.
We define $\cI_A$ to be the subspace of functions that do not depend on the configurations 
$\omega_{[N] \setminus A}$:
\begin{eqnarray*}
   \cI_A & := & \left\{ f \in {\Bbb R}^{\Omega_V} \; : \; \right. \\
         &    & \left. \mbox{$f(\omega_A,\omega_{V \setminus A}) =  
                f(\omega_A,\omega_{[N] \setminus A}')$ for all 
                $\omega_A \in \Omega_A$, and all $\omega_{[N] \setminus A},\omega_{[N] \setminus A}' 
                \in \Omega_{[N] \setminus A}$}\right\}.
\end{eqnarray*}
The orthogonal projection $\Pi_A$ 
onto this $|\Omega_A|$-dimensional space 
with respect to the canonical scalar product
\[
  \langle f,g\rangle \; := \; \sum_{\omega \in \Omega_V} f(\omega) g(\omega) 
  \qquad (f,g \in {\Bbb R}^{\Omega_V})
\]
 in ${\Bbb R}^{\Omega_V}$ is given by
\[
  \Pi_A(f)(\omega_A,\omega_{[N] \setminus A}) \; := \; \frac{1}{|\Omega_{[N] \setminus A}|}
  \sum_{\omega_{[N] \setminus A}'\in \Omega_{[N] \setminus A}} f(\omega_A,\omega_{[N] \setminus A}').
\]
In order to describe only the pure contributions of $A$ to a function $f$, we "subtract" the 
contributions from subsets $B \subsetneq A$. This leads to the 
$\prod_{i \in A}\big(|\Omega_i| - 1\big)$-dimensional subspace 
\[
  \widetilde{\cI}_A \; := \; 
  {\cI}_A \cap \left( \bigcap_{B \subsetneq A} {\cI_{B}}^\perp\right)
\]
and the orthogonal decomposition 
${\Bbb R}^{\Omega_V} = \bigoplus_{A\subseteq [N]}\widetilde{\cI}_A$. 
Denoting the orthogonal projections onto $\widetilde{\cI}_A$ by 
$\widetilde{\Pi}_A$ we thus have 
$\widetilde{\Pi}_A\widetilde{\Pi}_B=\delta_{A,B}\widetilde{\Pi}_A$ and
\begin{equation} \label{zerl}
     {\Pi}_A \; = \; \sum_{B \subseteq A} \widetilde{\Pi}_B, \qquad A \subseteq [N],
\end{equation}
and every vector $f$ has a unique representation as a sum of orthogonal vectors:
\[
     f \; = \; \sum_{A \subseteq [N]} \widetilde{\Pi}_A(f).
\]
The $f_A$ is called ({\em pure\/}) {\em interaction\/} among the units in $A$.
With the M\"obius inversion (\ref{zerl}) implies
\begin{eqnarray*}
   \widetilde{\Pi}_A(f) 
            & = & \sum_{B \subseteq A} (-1)^{|A \setminus B|} \Pi_B(f) \\
            & = & \sum_{B \subseteq A} (-1)^{|A \setminus B|}  
                  \frac{1}{|\Omega_{[N] \setminus B}|}
                  \sum_{\omega_{[N] \setminus B}'\in \Omega_{[N] \setminus B}} 
                  f(\omega_B,\omega_{[N] \setminus B}').   
\end{eqnarray*}
Now we construct exponential families associated with such interaction spaces. 
The most general construction is based on a set 
of subsets of $[N]$. Given such a set $\boldsymbol{A} \subseteq 2^{[N]}$, we define the corresponding
interaction space by 
\beq
   \widetilde{\cI}_{\boldsymbol{A}} \; := \; \bigoplus_{A \in \boldsymbol{A}}
   \widetilde{\cI}_A, 
\Leq{fat:A}
which generates the exponential family $\exp(\widetilde{\cI}_{\boldsymbol{A}})$. 
We want to apply this definition to the more specific situation of interactions with fixed {\em order} $k$.  
Therefore, we define
\[
   {\cI}^{(k)} \; := \; \widetilde{\cI}_{\{A \subseteq [N] \; : \; |A| \leq k\}}, \qquad 
   \mbox{and} \qquad 
   \widetilde{\cI}^{(k)} \; := \; \widetilde{\cI}_{\{A \subseteq [N] \; : \; |A| = k\}}.  
\]
We get the flag of vector spaces
\[
    {\Bbb R} \cong {\cI}^{(0)} \; \subsetneq \; {\cI}^{(1)} \; \subsetneq \; {\cI}^{(2)} \; 
    \cdots \; \subsetneq 
    {\cI}^{(N)} = {\Bbb R}^{\Omega_V},
\]
and the corresponding hierarchy of exponential families
\[
    \exp({\cI}^{(0)}) \; \subsetneq \; \exp({\cI}^{(1)}) \; \subsetneq \; \exp({\cI}^{(2)})
    \; \cdots \; \subsetneq 
    \exp({\cI}^{(N)}) = {\mathcal P}({\Omega_V}),
\]
Here, $\exp({\cI}^{(0)})$ contains exactly one element, namely the center of 
the simplex. 

The exponential family $\exp({\cI}^{(1)})$ is nothing but the exponential family ${\mathcal F}$ of 
factorizable distributions. Thus, the multi-information vanishes exactly 
on the topological closure of $\exp({\cI}^{(1)})$. 

Now we determine for a nonempty set $\cM(\Omega_1,\dots,\Omega_N)$ 
of maximizers the lowest order $k$ such that 
$\cM(\Omega_1,\dots,\Omega_N)$ is contained in the
topological closure of $\exp({\cI}^{(k)})$. The first possible candidate 
for this is given by $k= 2$. The following theorem states that this is also 
sufficient.
\medskip   

\begin{theorem}\label{thm:main} 
There exists an exponential family $\cF^*\subseteq \exp(\widetilde{\cI}^{(2)})$
of dimension $\dim(\cF^*)= (n_N-1) \sum_{i=1}^{N-1} (n_i-1)$ containing 
in its closure all global maximizers of the multi-information 
($\cM(\Omega_1,\dots,\Omega_N)\subset\overline{\cF^*}$).
\end{theorem}
\medskip

This theorem represents our main result which we already stated informally in Section~3. 
Note that compared with Theorem \ref{thm:starkered} for large $N$ Theorem \ref{thm:main}  leads to an exponential family 
$\cF^*$ of higher dimension. On the other hand, we still have 
an exponential (in $N$) codimension in the simplex 
$\overline{\cP}(\Omega_V)$. 

In addition to that, 
the exponential family of Theorem \ref{thm:main} represents a concrete model  
that appears in many applications in physics and biology. For instance, within the field of neural networks, the 
exponential family $\exp({\cI}^{(2)})$, which contains $\exp(\widetilde{\cI}^{(2)})$ as a subfamily, 
is known as the family of Boltzmann machines, \cite{AHS,AK,AKN}. 
Applied to this context, our result 
states that Boltzmann machines are able to generate all distributions that have globally maximal 
multi-information, and that their dimensionality ${N\choose 2}$ 
is not minimal for $N>2$.

\begin{examples} \label{remarksuff}$ $\\
{\em 
{\bf (1)} {\bf The Case of Two Units.} In this case, the hierarchy of interactions ends with $k=2$, 
because we have just two units. Thus the simplex ${\mathcal P}(\Omega_1 \times \Omega_2)$ is equal to 
the exponential family $\exp ({\mathcal I}^{(2)})$, which has dimension $n_1 n_2 - 1$. The codimension of 
the subfamily $\exp (\widetilde{\mathcal I}^{(2)})$ of Theorem \ref{thm:main} then is $n_1 + n_2 -2$. 
Applied to our example of two binary units from the introduction, we see that 
\[
   \dim(\exp (\widetilde{\mathcal I}^{(2)})) \; = \; 1
\]   
In Figure 1, we obtain this family by simply taking the convex combinations of the two maximizers:
\[
   \exp (\widetilde{\mathcal I}^{(2)}) \; = \; \left\{\frac{1-\lambda}{2} 
   \left( \delta_{(0,0)} + \delta_{(1,1)} \right) + \frac{\lambda}{2} 
   \left( \delta_{(1,0)} + \delta_{(0,1)} \right) \; : \; 0 < \lambda < 1 \right\}.
\]
{\bf (2)} {\bf The Case of $N$ Equal Units.} 
According to Theorem \ref{prop1} for $|\Omega_i|=n$ we have 
$|\cM(\Omega_1,\dots,\Omega_N)|=(n!)^{N-1}$ maximizers, which are, 
according to Theorem \ref{thm:main}, contained in the closure of an
exponential family $\cF^*$ of pure pair interactions,  with
\[\dim(\cF^*)=(N-1)(n-1)^2.\]
}
\end{examples}


\Section{\bf Proofs}
We fix the following notations:
For $V' \subset [N]$, 
$H_{V'}$ denotes the entropy of the random variable $X_{V'}$. Obviously $H_V = H$, and $H_{\{i\}} = H_i$. 
For two subsets $V',V'' \subset [N]$,    
$H_{(V'' \, | \, V')}$ is the conditional entropy of $X_{V''}$ given $X_{V'}$. For $V'= \{a_1,\dots,a_L\}$ and
$V''=\{b_1,\dots,b_M\}$ we also write $H_{(b_1,\dots,b_M \, | \, a_1,\dots,a_L)}$ instead of 
$H_{(V'' \, | \, V')} = H_{(\{b_1,\dots,b_M\} \, | \, \{a_1,\dots,a_L\})}$.
Now let $V_1,\ldots,V_r$ be a set of disjoint subsets of $[N] = \{1,\ldots,N\}$. 
The multi-information of these subsystems
is given by $I_{\{V_1,\ldots,V_r\}} = \sum_{j=1}^r H_{V_j} - H_{V_1 \uplus \dots \uplus V_r}$. 
In the case where the subsets of $[N]$ have cardinality one, we also write 
$I_{\{i_1,\dots,i_r\}}$ instead of $I_{\{\{i_1\},\dots,\{i_r\}\}}$. We obviously have $I_V = I$.\\

{\bf Proof of Lemma \ref{lemma1}.}\\
By the chain rule $H(X,Y)=H(X)+H(Y\mid X)$
\beqno
\lefteqn{   I_p(X_1,\dots,X_N) \;  =  \; 
   \sum_{i=1}^N H_p(X_i) \; - \; H_p(X_1,\dots,X_N)}&& \\
  & = &\sum_{i=1}^{N-1} H_p(X_i) - \big(H_p(X_1,\dots,X_N)-H_p(X_N)\big)\\
  & = &\sum_{i=1}^{N-1} H_p(X_i) - H_p(X_1,\dots,X_{N-1}\mid X_N)
   \leq \; \sum_{i=1}^{N-1} H_p(X_i)\leq \sum_{i=1}^{N-1} \ln(n_i),
\eeqno
proving the lemma.\hfill $\Box$\\

{\bf Proof of Theorem \ref{prop1}.}\\ 
If a probability distribution $p$ on $\Omega_V$
has the form (\ref{darstellung}) with a distribution $p^{(N)} \in \overline{\mathcal P}(\Omega_N)$
and surjective maps $\pi_i: \Omega_N \rightarrow \Omega_i$ that satisfy 
(\ref{surjektionen}), then $I(p) = \sum_{i=1}^{N-1} \ln(n_i)$:

\beqno   
\lefteqn{     I(p) = \sum_{i=1}^N H_i(p) - H(p)}&&   \\
        & = & \sum_{i=1}^N H_i(p) - H_N(p) 
         \underbrace{- H_{(1 | N)}(p) - H_{(2 | 1,N)}(p) - \dots 
              - H_{(N-1 |1,2,\dots,N-2,N)}(p)}_{=0} \\
        & = & \sum_{i=1}^{N-1} \ln(n_i).
\eeqno

Now we prove the opposite implication. Therefore we assume 
$I(p) = \sum_{i=1}^{N-1} \ln(n_i)$. This gives us
\begin{equation} \label{volleentropie}
   H_i(p) \; = \; \ln(n_i) \qquad  (i=1,\dots,N-1).
\end{equation}

Otherwise the existence of an $i_0 \in \{1,\dots,N-1\}$ with $H_{i_0}(p) < \ln(n_{i_0})$ would
imply the following contradiction 

\begin{eqnarray*}
   I(p) & = & \sum_{i = 1}^N H_i(p) - H(p)                                 \\
        & = & \sum_{i = 1}^{N-1} H_i(p) + H_N(p) 
              - \big( H_N(p) + H_{(1,\dots,N-1 \, | \, N)}(p) \big)        \\
        & \leq &   \sum_{i = 1 \atop i\not= i_0}^{N-1} H_i(p) + H_{i_0}(p) 
        \; <  \;   \sum_{i=1}^{N-1} \ln(n_i).
\end{eqnarray*}

 From (\ref{volleentropie}) we have

\begin{equation} \label{pkagleichp}
  H(p) \; = \; \sum_{i=1}^N H_i(p) - I(p) \; = \; \left(\sum_{i = 1}^{N-1} \ln(n_i) + H_N(p)\right) - 
               \sum_{i=1}^{N-1} \ln(n_i)  \; = \; H_N(p).
\end{equation}

Now we set $p^{(N)} := p_N$, and define a Markov kernel 
$K: (\Omega_1\times\cdots\times\Omega_{N-1}) \times \Omega_N \rightarrow [0,1]$ by

\[
  K(\omega_1,\dots,\omega_{N-1} \, | \, \omega_N) \; := \; 
  \left\{ 
     \begin{array}{c@{,\quad}l}
        \frac{p(\omega_1,\dots,\omega_N)}{p_N(\omega_N)} & \mbox{if $p_N(\omega_N) > 0$} \\
        \frac{1}{n_1\cdots n_{N-1}} & \mbox{if $p_N(\omega_N) = 0$}
     \end{array}
  \right..
\]

In these definitions we get

\beqno
\lefteqn{H(p) - H_N(p)} \\
& = & \sum_{\omega_N \in \Omega_N \atop p_N(\omega_N) > 0} 
      p_N(\omega_N) 
      \Bigg( \ln p_N(\omega_N) - \\
&   &    \sum_{(\omega_1,\dots,\omega_{N-1}) \in \atop \Omega_1\times\cdots\times\Omega_{N-1}}
         K(\omega_1,\dots,\omega_{N-1} \, | \, \omega_N) 
         \ln \Big( p_N(\omega_N) \, K(\omega_1,\dots,\omega_{N-1} \, | \, \omega_N) \Big)
      \Bigg) \\
& = & \sum_{\omega_N \in \Omega_N \atop p_N(\omega_N) > 0} 
      p_N(\omega_N) \, H\big(K(\cdot \, | \, \omega_N)\big) \; \geq \; 0.
\eeqno

 From (\ref{pkagleichp}) this implies $H\big(K(\cdot \, | \, \omega_N)\big) = 0$ for all $\omega_N$
with $p_N(\omega_N) > 0$. This implies the existence of maps 
$\pi_i: \Omega_N \rightarrow \Omega_i$ with 

\[
   p(\omega_1,\dots,\omega_N) \; = \; p^{(N)}(\omega_N) \prod_{i=1}^{N-1} 
   \delta_{\omega_i, \pi_i(\omega_N)}.
\] 

Because of $H_i(p) = \ln(n_i)$ for all $i \in \{1, \dots,N-1\}$, these maps 
must be surjective. \hfill $\Box$ \\[2mm]

{\bf Proof of Theorem \ref{thm:4.4}.} \\
{\bf Proof that $\cM(\Omega_1,\ldots,\Omega_N)\neq\es$ if $n_N\ge n_{\min}$:}\\ 
For $m\in\bN$ set $T_m:=\l\{\frac{i}{m}\ : \  i\in[m]\ri\}$. We claim that the
cardinality of
\[T_\Omega:=\bigcup_{i\in[N-1]}T_{n_i}\] 
is given by
$|T_\Omega|=n_{\min}$.
This follows by the inclusion--exclusion principle if
\beq
\l|\bigcap_{i\in A}T_{n_i}\ri|={\rm GCD}(n_A) \qquad (A\in W),
\Leq{A}
since
\[\l|\bigcup_{i\in[N-1]}T_{n_i}\ri|=\sum_{A\in W}(-1)^{|A|-1}\l|\bigcap_{i\in A}
T_{n_i}\ri|.\]
To prove (\ref{A}), we set $m_A:={\rm GCD}(n_A)$ and note that $T_{n_i}\supseteq T_{
m_A}\ \ (i\in A)$. Thus $\l|\bigcap_{i\in A}T_{n_i}\ri|\geq\l|T_{m_A}\ri|=
m_A$.

To show the converse inequality $\l|\bigcap_{i\in A}T_{n_i}\ri|\leq
m_A$ we note that for some $\tilde{m}\in\bN$ we have $\bigcap_{i\in A}T_{n_i}
=T_{\tilde{m}}$. 
Thus for all $i\in A$ there exist $\ell_i\in[n_i]$ with 
$\frac{\ell_i}{n_i}=\frac{1}{\tilde{m}}=\min(T_{\tilde{m}})$, 
or $n_i=\ell_i\tilde{m}$. 
Thus $\tilde{m}$ divides all
$n_i\ \ (i\in A)$ and -- being the largest such integer -- equals 
$m_A={\rm GCD}(n_A)$.

Now we write $T_{\Omega_V}$ in the form $\{d_1,\ldots,d_{n_{\min}}\}$ and set
$d_0:=0$, with ordering $d_i>d_{i-1}\ \ (i\in[n_{\min}])$.
The map 
\[\Phi:T_{\Omega_V}\to\Omega_V\qmbox{,} \Phi(d_j)_i:=\l\{\begin{array}{ccc}
\lceil d_jn_i\rceil&,&i\in[N-1]\\ j&,&i=N\end{array}\ri.\] 
is well defined,
since $\lceil d_jn_i\rceil\in[n_i]\ \ (i\in[N-1])$, and by our assumption 
$n_N\geq n_{\min}$ which implies $j\in[n_N]$.
The function 
\beq
p:\Omega_V\to\bR\qmbox{,} p:=\sum_{j=1}^{n_{\min}}(d_j-d_{j-1})\delta_
{\Phi(d_j)}
\Leq{p:c}
is a probability distribution since $d_j-d_{j-1}>0$ and 
\[\sum_{j=1}^{n_{\min}}(d_j-d_{j-1})=d_{n_{\min}}-d_0=1.\]
For all $i\in[N-1]$ and $\ell\in[n_i]$ the $i$th marginal probability equals
\beqno
p_i(\ell)&=&
\sum_{\omega\in \times_{j\in[N]\backslash\{i\}}\ n_j}p(\ell,\omega)=
\sum_{\stackrel{\omega_N\in[n_{\min}]}{\lceil d_{\omega_N}n_i\rceil=\ell}}
(d_{\omega_N}-d_{\omega_N-1})=\\
&=&\sum_{j:d_j\in\l(\frac{\ell-1}{n_i},\frac{\ell}{n_i}\ri]}(d_j-d_{j-1})=
\frac{\ell}{n_i}-\frac{\ell-1}{n_i}=\frac{1}{n_i}.
\eeqno
We thus meet the condition of Theorem 3.2 showing that $p\in\cM(\Omega_1,\ldots,\Omega_N)$.
\hfill
\vspace{2mm}

{\bf Proof that $\cM(\Omega_1,\ldots,\Omega_N)=\es$ if $n_N< n_{\min}$:} 
\begin{itemize}
\item
The statement is trivial for $N=2$ (remember that we assume $n_{i+1}\geq
n_i$). Assume now that it is proven for all product spaces of at most
$N\in\bN$ units. Then for a probability
distribution $p\in\cM(\Omega_1,\ldots,\Omega_{N+1})$ consider its marginal
$\tp\in\ov{\cP}(\Omega_{[N]})$.
\\
We associate to $\tp$ a $N$--partite graph $(\tilde{V},E)$ whose vertex set
is the disjoint union
$\tilde{V}:=\dot{\bigcup}_{i=1}^N\Omega_i$.
To every $\omega=(\omega_1,\ldots,\omega_N)\in\supp(\tp)\subseteq\Omega_{[N]}$
belongs the complete graph on the vertex set 
$\{\omega_1,\ldots,\omega_N\}\subset \tilde{V}$ with edge set 
$G_\omega:=\big\{\{\omega_i,\omega_j\}\subset \tilde{V}\mid1\leq i<j\leq N\big\}$
on the $N$ vertices $\omega_1,\ldots,\omega_N$.
Then the edge set 
\[E:=\bigcup_{\omega\in\supp(\tp)}G_\omega\]
on $\tilde{V}$ is indeed $N$--partite.
By the strict positivity (\ref{surjektionen}) of the $p$--marginals
no vertex $v\in \tilde{V}$ is isolated.
\item

Every edge set $G_\omega\subseteq E$ is contained in the induced subgraph of
exactly one connected component $C\subseteq \tilde{V}$ of the graph $(\tilde{V},E)$.
We attribute to $G_\omega$ the weight $\tp(\omega)$, and to a
connected component $C$ of the graph $(\tilde{V},E)$ 
the sum of the weights of the $G_\omega$ contained in it.
\begin{figure}
\begin{picture}(0,0)%
\includegraphics{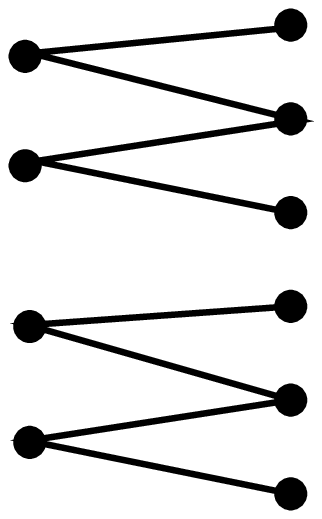}%
\end{picture}%
\setlength{\unitlength}{3947sp}%
\begingroup\makeatletter\ifx\SetFigFont\undefined%
\gdef\SetFigFont#1#2#3#4#5{%
  \reset@font\fontsize{#1}{#2pt}%
  \fontfamily{#3}\fontseries{#4}\fontshape{#5}%
  \selectfont}%
\fi\endgroup%
\begin{picture}(2547,2991)(151,-2219)
\put(400,-2000){\makebox(0,0)[lb]{\smash{{\SetFigFont{12}{14.4}{\familydefault}{\mddefault}{\updefault}{$\Omega_1$}%
}}}}
\put(1700,-2000){\makebox(0,0)[lb]{\smash{{\SetFigFont{12}{14.4}{\familydefault}{\mddefault}{\updefault}{$\Omega_2$}%
}}}}
\put(1050,-2000){\makebox(0,0)[lb]{\smash{{\SetFigFont{12}{14.4}{\familydefault}{\mddefault}{\updefault}{$E$}%
}}}}
\put(2206,269){\makebox(0,0)[lb]{\smash{{\SetFigFont{12}{14.4}{\rmdefault}{\mddefault}{\updefault}{$C_1$}%
}}}}
\put(2176,-1111){\makebox(0,0)[lb]{\smash{{\SetFigFont{12}{14.4}{\rmdefault}{\mddefault}{\updefault}{$C_2$}%
}}}}
\end{picture}%
\caption{A bipartite graph $(\tilde{V},G)$ for $N+1=3$ units, with $|\Omega_1|=4$, 
$|\Omega_1|=6$, $|\Omega_3|\ge 8$ and a maximizer 
$p\in \cM(\Omega_1,\Omega_2,\Omega_{3})$. $(\tilde{V},G)$ has the 
two components $C_1,C_2$.}
\end{figure}
These weights $w(C)$ of the connected components $C$
are not arbitrary numbers in $(0,1]$. 
Instead, we know from Theorem \ref{prop1} that the marginal distributions 
$p_i:\Omega_i\to[0,1]$ of $p$ (and thus of $\tp$, too) have the Laplace form
\[p_i(\omega_i)=\frac{1}{n_i} \qquad (i\in[N],\ \omega_i\in\Omega_i).\]
Therefore $w(C)$ is simultaneously an integer multiple of $1/n_i\ \ (i\in [N])$
and thus an integer multiple of ${\rm GCD}(n_{[N]})$.
This implies the upper bound ${\rm GCD}(n_{[N]})$ for the number of connected
components $C$ of the $N$--partite graph $(\tilde{V},E)$.
\item
For the case of $N+1=3$ units this already suffices to show the bound $n_3\geq
n_{\min}=n_1+n_2-\GCD(n_1,n_2)$.
In this case the complete graphs are of cardinality $|G_\omega|=(N-1)!=1$ so that
$|E|=|\supp(\tp)|$.
\\
In general a graph on a vertex set of $v\in\bN$ vertices with $e\in\bN_0$ edges
has at least $\max(v-e,1)$ connected components.
In the case at hand $v=n_1+n_2$, and there are at most $\GCD(n_1,n_2)$ connected
components. So
\beqno
n_3&\geq& |{\rm supp} ({p})|\geq|{\rm supp} (\tilde{p})| = |E|=e\\
&\geq& v-c=(n_1+n_2)-{\rm GCD}(n_1,n_2)=n_{\min}.
\eeqno
\item
For arbitrary $N+1>3$ this argument must be modified, since then $|G_\omega|=(N-1)!>1$.
\\
First of all we can substitute $G_\omega$ by any spannning tree
$T_\omega\subset G_\omega$, and 
still the connected components $C'$ of $(\tilde{V},E')$ with $E':=\bigcup_
{\omega\in\supp(\tp)}T_\omega$ coincide with the connected components $C$ of $(\tilde{V},E)$.
Each of these spanning trees has only $|T_\omega|=N-1$ edges. 
However in general $E'$, too is not a disjoint union of the $T_\omega$.
\\
We thus decompose the set $\supp(\tp)$ into a disjoint union 
\beq 
\supp(\tp)=\bigcup_{k=1}^NA_k,
\Leq{parti}
beginning with an arbitrarily chosen set $A_N$ of representatives $\omega\in C$ 
of the connected
components $C\subseteq\Omega_{[N]}$. 
The estimate on the number of these components implies 
$|A_N|\geq\GCD(n_{[N]})$, and for $\omega\neq \omega'\in A_N$ the edge sets 
$G_\omega$ and $G_{\omega'}$ are disjoint.
\\
Next we arrange the elements $\omega'\in C$ of the connected component 
$C$ containing $\omega\in A_N$ in the form of a spanning tree, with 
$G_{\omega'}\cap G_{\omega''}\neq\es$ for $\{\omega',\omega''\}$ 
being an edge of that tree. For $\omega'=(\omega_1',\ldots,\omega_N')\in C$ 
of distance $d(\omega')$ from 
$\omega\in A_N$ we put $\omega'\in A_k$ if there are exactly 
$k$ indices $i\in[N]$ with $\omega_i'$ not
being equal to any $\omega_i''$ for 
$\omega''=(\omega_1'',\ldots,\omega_N'')$ with $d(\omega'')<d(\omega')$.
This indeed gives a partition of the form (\ref{parti}).
\\
Then by our induction hypothesis
\beq
|A_k|\geq\sum_{\stackrel{B\subseteq[N]}{|B|\geq k}}(-1)^{|B|-k}{|B|\choose k}\GCD(n_B)
\qquad (k=1,\ldots,N).
\Leq{A:k}
Namely for $k=N$ (\ref{A:k}) reduces to $|A_N|\geq\GCD(n_{[N]})$ which has been shown 
to be true. So if (\ref{A:k}) would not hold,
for the smallest 
$k<N$ violating (\ref{A:k}), we would find
a $B\subseteq[N]$ of cardinality $|B|=k<N$, whose marginal distribution $p_B$
has support of cardinality 
$\hat{n}_{k+1}:= |\supp(p_B)| < n_{\min}(B)=
\sum_{\stackrel{\tilde{B}\subseteq B}{\tilde{B}\neq\es}}
(-1)^{|\tilde{B}|-1}\GCD(n_{\tilde{B}})$, see
(\ref{sumsum}) below.

But this would contradict our induction assumption, since then the system
$\hat{\Omega}:=\big(\times_{i\in B} [n_i]\big)\times[\hat{n}_{k+1}]$ 
would have the optimizing probability distribution 
\[\hat{p}:\hat{\Omega}\ar[0,1]\qmbox{,} 
\hat{p}(\omega_B,l):=\delta_{e(l),\omega_B}\ p_B(\omega_B)\]
for some bijection $e:[\hat{n}_{k+1}]\ar \supp(p_B)$, 
but yet not meet the criterium $\hat{n}_{k+1}\geq  n_{\min}(B)$.
\\[2mm]
Summing the cardinalities (\ref{A:k}), we obtain
\beqn
|\supp(\tp)|&=&\sum_{k=1}^N|A_k|\geq\sum_{k=1}^{[N]}\sum_{\stackrel{B\subseteq[N]}
{|B|\geq k}}(-1)^{|B|-k}{|B|\choose k}\GCD(n_B)\nonumber\\
&=&\sum_{\stackrel{B\subseteq[N]}{B\neq\es}}(-1)^{|B|}\GCD(n_B)\sum_{k=1}^{|B|}(-1)^k
{|B|\choose k}\nonumber\\
&=&\sum_{\stackrel{B\subseteq[N]}{B\neq\es}}(-1)^{|B|-1}\GCD(n_B)=n_{\min},
\label{sumsum}
\eeqn
which is the induction step.
\hfill $\Box$
\end{itemize}
{\bf Proof of Lemma \ref{lem:poset}.}\\
If $n_1=n_2$ then the maps $\pi\in {\mathcal S}$ are isomorphisms 
$\pi:\Omega_2\to\Omega_1$, so that $\sigma\preceq\pi$ only for $\sigma=\pi$.
Thus in that case ${\mathcal S}$ is connected iff $|{\mathcal S}|=1$, i.e.\ $n_1=n_2=1$. This contradicts 
our assumption $n_1,n_2 \geq 2$.

If $n_2>n_1$ and $|\pi^{-1}(\omega_1)|>1$ for $\pi\in {\mathcal S}$ and some
$\omega_1\in\Omega_1$, say $\pi(\omega'_2)=\omega_1$, then 
$\sigma\preceq\pi$ for 
\[
  \sigma\in {\mathcal S}, \qquad \sigma(\omega_2):= 
  \l\{
    \begin{array}{c@{,\quad}l}  
      \pi(\omega_2) & \mbox{if $\omega_2 \neq \omega'_2$}  \\ 
      0             & \mbox{if $\omega_2 = \omega'_2$}
    \end{array} 
  \ri..
\]
So we need only show that any $\pi',\pi''\in S$ which are injective onto $\Omega_1$
are indeed connected.

\begin{enumerate}
\item
In the first step we move $\pi'$ along the poset graph in order to decrease
the cardinality of the symmetric difference $(\pi')^{-1}(0)\Delta(\pi'')^{-1}
(0)$. So we assume that there exist
\[\omega'\in(\pi')^{-1}(0)\backslash(\pi'')^{-1}(0) \qmbox{and}
\omega''\in(\pi'')^{-1}(0)\backslash(\pi')^{-1}(0)\]
and set
\[
  \pi \in {\mathcal S}, \qquad \pi(\omega) := 
  \l\{
    \begin{array}{c@{,\quad}l}
      0               & \mbox{if $\omega = \omega''$} \\
      \pi'(\omega'')  & \mbox{if $\omega=\omega'$} \\
      \pi'(\omega)    & \mbox{otherwise}. 
    \end{array}
  \ri.
\]
Both $\pi'$ and $\pi$ are covered by
\[
  \rho\in {\mathcal S}, \qquad 
  \rho(\omega) := 
  \l\{
    \begin{array}{c@{,\quad}l}
      \pi'(\omega'')   & \mbox{if $\omega=\omega'$} \\
      \pi'(\omega)     & \mbox{otherwise,}
    \end{array}
  \ri.
\]
and
\[
  |\pi^{-1}(0)\Delta(\pi'')^{-1}(0)| \; = \; |(\pi')^{-1}(0)\Delta(\pi'')^{-1}(0)|-2.
\]
By iterating the argument we can assume w.l.o.g.\ that $(\pi')^{-1}(0) = 
(\pi'')^{-1}(0)$.
\item
In fact it is sufficient to treat the case where the permutation
\[
  \pi''\circ(\pi')^{-1}\mid_{\Omega_1}:\Omega_1\to\Omega_1
\]
is a transposition, as the transpositions generate the symmetric group. So there
exist $\omega^I\neq\omega^{II}\in\Omega_2$ with
\[
  \pi''(\omega) = 
  \l\{
    \begin{array}{c@{,\quad}l}
      \pi'(\omega^I)    & \mbox{if $\omega = \omega^{II}$} \\
      \pi'(\omega^{II}) & \mbox{if $\omega = \omega^I$}    \\
      \pi'(\omega)      & \mbox{otherwise,}
    \end{array}
  \ri.
\]
and we choose $\hat{\omega}\in\Omega_2$ so that $\pi'(\hat{\omega})=\pi''
(\hat{\omega})=0$.

Defining $\rho,\rho''\in {\mathcal S}$ by
\[
  \rho'(\omega) := 
    \l\{
      \begin{array}{c@{,\quad}l}
        \pi'(\omega^{II})  & \mbox{if $\omega = \hat{\omega}$} \\
        0                  & \mbox{if $\omega = \omega^{II}$}  \\
        \pi'(\omega)       & \mbox{otherwise} 
      \end{array}
    \ri.
\qmbox{resp.}
  \rho''(\omega) := 
    \l\{
      \begin{array}{c@{,\quad}l}
        \pi''(\omega^I)   & \mbox{if $\omega=\hat{\omega}$} \\
        0                 & \mbox{if $\omega=\omega^{I}$}   \\
        \pi''(\omega)     & \mbox{otherwise,} 
      \end{array}
    \ri.
\]
$\pi'$ and $\rho'$ are covered by $\sigma' \in {\mathcal S}$ and similarly $\pi''$ and 
$\rho''$ are covered by $\sigma''\in {\mathcal S}$ with
\[
  \sigma''(\omega) := 
    \l\{
      \begin{array}{c@{,\quad}l}
        \pi'(\omega^{II})   &  \mbox{if $\omega=\hat{\omega}$}       \\
        \pi'(\omega)        &  \mbox{otherwise} 
      \end{array}
    \ri.
\qmbox{resp.}
  \sigma''(\omega) := 
    \l\{
      \begin{array}{c@{,\quad}l}
        \pi''(\omega^I)  & \mbox{if $\omega = \hat{\omega}$}  \\
        \pi''(\omega)    & \mbox{otherwise.} 
      \end{array}
    \ri.
\]
Now as $\pi'(\omega^{II}) = \pi''(\omega^I)$, both $\rho'$ and $\rho''$ are
covered by
\[
  \tau \in {\mathcal S}, \qquad
    \tau(\omega) := 
    \l\{
      \begin{array}{c@{,\quad}l}
        \pi'(\omega^{II}) & \mbox{if $\omega = \hat{\omega}$} \\
        \pi'(\omega^I)    & \mbox{if $\omega = \omega^{II}$}  \\
        \pi'(\omega)      & \mbox{otherwise.}
      \end{array}
    \ri.
\]
This shows that the poset graph is connected.
\hfill $\Box$
\end{enumerate}
\bigskip

{\bf Proof of Theorem \ref{thm:two:units}.}\\ 
To simplify notation, we set ${\mathcal M} := {\mathcal M}(\Omega_1,\Omega_2)$, and 
${\mathcal M}_\pi := {\mathcal M}_\pi(\Omega_1,\Omega_2)$ for $\pi \in {\mathcal S}$.\\
{\bf (1)} 
We have $\cM_\pi\subset \cM$ 
since for the elements of $\cM_\pi$ 
the characterisation of Theorem \ref{prop1} hold true. Furthermore for $\sigma,
\pi\in {\mathcal S}$ with $\sigma\neq\pi$ there exists $(\omega_2,\omega_1)\in{\rm graph}(\pi)$
with $(\omega_2,\omega_1)\not\in{\rm graph}(\sigma)$ or vice versa. Thus for
$p\in \cM_\pi$ we have $p(\omega_1,\omega_2)>0$ but for $p\in \cM_\sigma$ we
have $p(\omega_1,\omega_2)=0$ showing that $\cM_\pi\cap \cM_\sigma=\emptyset$. \\
Finally for $p\in \cM$ by Theorem \ref{prop1} 
there exists a surjective map $\tilde{\pi}
:\Omega_2\to\Omega_1$ with $p(\omega_1,\omega_2)=0$ whenever $\tilde{\pi}
(\omega_2)\neq\omega_1$. Given $\tilde{\pi}$, we construct $\pi\in {\mathcal S}$ by
setting
\[
  \pi(\omega_2) := 
    \left\{
      \begin{array}{c@{,\quad}l}
        \tilde{\pi}(\omega_2)  & \mbox{if $p(\tilde{\pi}(\omega_2),\omega_2) > 0$} \\
        0                      & \mbox{if $p(\tilde{\pi}(\omega_2),\omega_2) = 0$}.
      \end{array}
    \right.
\]
As by Theorem \ref{prop1} we have $\sum_{\omega_2\in\tilde{\pi}^{-1}(\omega_1)}
p(\omega_1,\omega_2)=\frac{1}{n_1}>0$, the function $\pi:\Omega_2\to\Omega_1^*$
so constructed has the property $\pi(\Omega_2)\supset\Omega_1$ making it an 
element of $S$. \\
{\bf (2)}
Given $\omega_1\in\Omega_1$, the simplex of $|\pi^{-1}(\omega_1)|$ numbers 
$p(\omega_1,\omega_2)>0$ with $\omega_2\in\pi^{-1}(\omega_1)$ meeting 
$\sum_{\omega_2\in\pi^{-1}(\omega_1)}p(\omega_1,\omega_2) = \frac{1}{n_1}$ has 
dimension $|\pi^{-1}(\omega_1)|-1$, implying the formula for $\dim \cM_\pi$.\\
If $\dim \cM_\pi = l-n_1$, the surjective map $\hat{\pi}:\hat{\Omega}_2\to
\Omega_1$ with $\hat{\Omega}_2 := \pi^{-1}(\Omega_1)\subset\Omega_2$ and
$\hat{\pi} := \pi\mid_{\hat{\Omega}_2}$ is defined on a subset $\hat{\Omega}_2
\subset\Omega_2$ of size $l$. There are precisely ${n_2\choose l}$ such
subsets, and there are precisely $n_1!S_{l,n_1}$ such surjective maps from
$\hat{\Omega}_2$ onto $\Omega_1$, see Aigner \cite{Ai}, Chapter 3.1. \\
{\bf (3)}
If $n_1=n_2$ then ${\mathcal S}$ coincides with the set of bijections $\pi:\Omega_2\to
\Omega_1$, and $|\cM_\pi|=1$. Thus in this case $\cM$ is not connected for
$n_1 \geq 2$. If, however $n_2>n_1$, the poset ${\mathcal S}$, seen as a graph, is connected.\\
The topological closure of $\cM_\pi$ is given by
\[
\overline{\cM}_\pi = \left\{p\in\overline{\mathcal{P}}(\Omega_1\times\Omega_2)
\; : \;  \sum_{\omega_2\in\pi^{-1}(\omega_1)}p(\omega_1,\omega_2) = \frac{1}{n_1},
p(\omega_1,\omega_2)=0\ \mbox{if}\ \pi(\omega_2)\neq\omega_1\right\}.
\]
Thus $\overline{\cM}_\pi = {\biguplus}_{\sigma\preceq\pi}\cM_\sigma$.
\hfill $\Box$
\bigskip

{\bf Proof of Corollary \ref{thm:same:card}.}\\
All statements directly follow from Theorem \ref{prop1} . \hfill $\Box$ \\[2mm]

{\bf Proof of Theorem \ref{thm:starkered}.}\\ 
We choose a map 
$\phi = (\phi_1,\dots,\phi_n): \; \Omega_V \; \rightarrow \; {\Bbb R}^n$ such that the points 
$\phi(\omega)$, $\omega \in \Omega_V$, are in general position; that is, each $k$ 
elements of $\phi(\Omega_V)$ 
with $k \leq n+1$ are affinely independent. This property guarantees that 
for each set $\Sigma \subset \Omega_V$, 
$|\Sigma| = n$, there exist real numbers $a_1,\dots,a_n,b$ such that
\begin{equation} \label{sigmasupport}
   \left\{ \omega \in \Omega_V \; : \; \sum_{i = 1}^n a_i \, \phi_i(\omega) = 
   b \right\} \; = \; \Sigma
\end{equation}
holds. We consider the exponential family ${\mathcal G}^\ast$ 
that is generated by $c$ and 
\[
   \phi_1,\dots,\phi_n\qmbox{,}\phi_i \, \phi_j\quad
   (1 \leq i \leq j \leq n). 
\]
We have 
\[
   \dim {\mathcal G}^\ast \; \leq  \; \frac{n^2 + 3n}{2}. 
\]
Now let $p$ be an element of $\cM(N \times n)$. 
 From Theorem \ref{thm:same:card} we know that $|{\rm supp}\, p | \; = \; n$. 
We prove that there exists a sequence 
in ${\mathcal G}^\ast$ that converges to $p$. 
We choose a sequence $\beta_m \uparrow \infty$ 
and real numbers 
$a_1,\dots,a_n,b$ satisfying (\ref{sigmasupport}) with $\Sigma = {\rm supp} \, p$. Then with
\[
   E^{(m)} \; := \; - \beta_m \left( \sum_{i=1}^n a_i \, \phi_i \; - \; b\right)^2,
\]
the sequence
\[
  \frac{\exp E^{(m)}}{\sum_{\omega' \in \Omega_V} \exp E^{(m)}(\omega')} \; \in \; {\mathcal G}^\ast  
\]
converges to $p$. \hfill $\Box$ \\[2mm]
{\bf Proof of Theorem \ref{thm:main}.}\\
Using def.\ (\ref{fat:A}), we consider for 
$\boldsymbol{A}:=\{\{1,N\},\{2,N\},\ldots,\{N-1,N\}\}\subset 2^{[N]}$
the linear subspace 
\[\widetilde{\cI}_{\boldsymbol{A}}\subset \widehat{\cI}_2\] 
of pure pair interactions of the $N$th unit with all other units.
The exponential family 
$\cF^*:= \exp(\widetilde{\cI}_{\boldsymbol{A}})\subset \cP(\Omega_V)$ 
is of dimension 
\[\dim(\cF^*)= (n_N-1)\sum_{i=1}^{N-1}(n_i-1),\]
as asserted in Theorem \ref{thm:main}.

Given a maximizer $p\in\cM(\Omega_1,\ldots,\Omega_N)$, we
now construct a sequence of probability distributions
\[q^{(m)}:= \exp(\tilde{f}^{(m)}) \in \cF^*\qquad (m\in \bN)\]
and show that
$\lim_{m\ar\infty}q^{(m)} = p$.

Here the functions $\tilde{f}^{(m)} \in \widetilde{\cI}_{\boldsymbol{A}}$ 
are defined as the orthogonal projections onto $\widetilde{\cI}_{\boldsymbol{A}}$
of ${f}^{(m)}\in {\cI}^{(2)}$
\[{f}^{(m)}(\omega):= 
\prod_{i=1}^{N-1} \delta_{\omega_i,\pi_i(\omega_N)}\ 
\frac{m+\ln\left(p^{(N)}(\omega_N)+1/m\right)}{N-1}\qquad (\omega\in\Omega_V) .\]
For $\omega,\omega'\in{\rm supp}(p)$
\begin{eqnarray*}
\frac{q^{(m)}(\omega)}{q^{(m)}(\omega')}
& = &\exp\left(\left[m+\ln\left(p^{(N)}(\omega_N)+\frac{1}{m}\right)\right]
\!\!-\!\!
\left[m+\ln\left(p^{(N)}(\omega_N')+\frac{1}{m}\right)\right]\right)\\
&=& \frac{p^{(N)}(\omega_N)+\frac{1}{m}}
{p^{(N)}(\omega_N')+\frac{1}{m}}\ \stackrel{m\ar\infty}{\longrightarrow}\ 
\frac{p^{(N)}(\omega_N)}{p^{(N)}(\omega_N')}
\end{eqnarray*}
in accordance with (\ref{darstellung}).

On the other hand if $\omega'\in{\rm supp}(p)$ but $\omega\not\in{\rm supp}(p)$, 
then there is an $i\in\{1,\ldots,N-1\}$ with $\omega_N\neq\pi_i^{-1}(\omega_i)$ or
$p^{(N)}(\omega_N)=0$.
In both cases
\[\lim_{m\to\infty}\frac{q^{(m)}(\omega)}{q^{(m)}(\omega')} = 0,\]
again in accordance with (\ref{darstellung}). As the $p^{(m)}$ are probability
distributions, we have shown that $\lim_{m\ar\infty}q^{(m)} = p$.
\hfill $\Box$ \\[2mm]

\bibliographystyle{amsalpha}

{\mi
\begin{flushright}
\begin{minipage}[]{124mm}
{Nihat Ay\\ Max Planck Institute for Mathematics in the Sciences,\\ Inselstr. 22, 
D 04103 Leipzig, nay@mis.mpg.de\\

Andreas Knauf\\ Mathematisches Institut,   
 Friedrich-Alexander-Universit\"at Erlangen-N\"urnberg, 
 Bismarckstr. 1 1/2,
D-91054 Erlangen, Germany,   
knauf@mi.uni-erlangen.de}
\end{minipage}
\end{flushright}
}

\end{document}